# Exploring Opportunistic Routing for Remote Sea Emergencies

Cleon Liew  
University of Nottingham  
psycl8@nottingham.ac.uk

Milena Radenkovic  
University of Nottingham  
milena.radenkovic@nottingham.ac.uk

*This paper explores the Opportunistic Routing Protocols in the context of remote sea emergency scenarios, using the MH370 plane crash as a case study (OppNetMH370). We studied the likelihood of successful transmissions of emergency messages to response services where communication methods are inadequate in remote sea areas. The study focuses on two opportunistic routing protocols, where their performances are evaluated based on key metrics including average latency and delivery probability. Our study reveals the challenges associated with the current communication technology in remote areas and proposes potential enhancements for future simulations. The findings contribute to understanding the limitations of existing communication strategies in remote sea areas and offers insights on the future development and improvements to the routing protocols.*(Abstract)

## I. INTRODUCTION

The disappearance of the Malaysia Airlines Flight MH370 on March 8, 2014 [2], is one of the most perplexing mysteries in aviation history. Despite the extensive searching, the aircraft's location has remained unknown for almost a decade [3]. This highlights the complexities of remote area search and rescue operations, and raises questions on the technology and capabilities of modern communication networks in emergency scenarios.

Considering the circumstances of the flight MH370, we assume that, in the event of a crash, some mobile nodes capable of wireless communication, such as electronic devices from passengers and crew, as well as potential debris, should have been dispersed from the crash site. The families of the passengers reported that phones continued to ring for up to four days after the plane went missing [4]. While wireless analysts suggest that the ringing tone could be a deliberate psychological mechanism to keep callers on the line [4], if the phones were indeed still active, it could signify a significant difference. In such a scenario, these active phones could potentially function as nodes capable of forming Mobile Ad-Hoc Networks (MANETs), Vehicular Ad-Hoc Networks (VANETs), or opportunistic networks.

The absence of effective communication or distress signals during the critical period prompts an exploration into the viability of leveraging Opportunistic networks for emergency communication in remote areas. The objective of this research is to simulate the performance attributes of opportunistic routing protocols in scenarios similar to the MH370 incident. Through the utilization of the ONE simulator, the aim is to recreate the probable post-crash conditions, where mobile devices or debris act as heterogenous network nodes, attempting to establish connections with sea vessels and emergency services.

## II. BACKGROUND

The disappearance of the plane in the scenario highlighted the challenges of remote emergency communication, where traditional networks are likely unavailable. Unlike many forwarding based protocols, where one message will only be carried by one node at a time, well-established replication based protocols will be the most suitable, as it aligns with the dynamic and sparse nature of the communication network. In this report, both Epidemic and MaxProp Protocol has been selected to be used in the simulation. Epidemic protocol is chosen as it is simple and resilient in situations with intermittent connectivity, while MaxProp protocol has been selected as it is efficient in utilising contact opportunities for message delivery. These two protocols align with the potential scenario of a remote plane crash, where various devices may act as nodes in a resource-constrained environment. While similar to one another, they have distinct characteristics and offer unique advantages in the context of the simulation of a remote plane crash.

## III. COMPLEX INFORMATION GATHERING AND SCENARIO DESIGN OF OPPNETMH370

MH370 impacted the ocean 1,933km due west of Perth at 33.177°S 95.300°E [5].

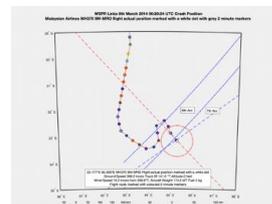 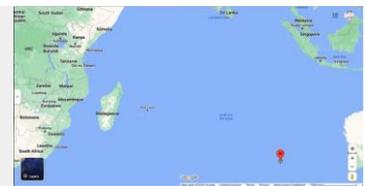

**Fig 1: Probable Crash Location**  **Fig 2: Probable Crash Location on Google Maps**

According to the live tracking AIS maps of ships current position, the density map of the Indian Ocean can be produced [6] as seen in Fig 3. Furthermore, ships can be tracked live as seen in Fig 4 [7].

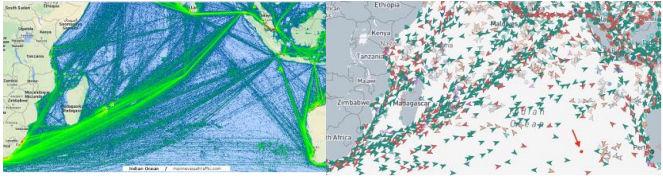

| **Fig 3: Indian Ocean Traffic Density** | **Fig 4: Screenshot of Live Ship Traffic** |

As the plane had went missing from the radar around MYT 2:40am [8] 2014, in order to estimate the sea traffic at that time, Fig 4 is a screenshot was taken at GMT 7:00pm (MYT 3:00am) on 7 Nov 2023. It also contains a red dot and an arrow pointing to it, indicating the probable crash location from Fig 1 and 2. In order to estimate the probable locations of other static nodes in the ocean, we were able to obtain the map of the Ocean moors from Data Buoy Cooperation Panel [9]. This is shown in Fig 5.

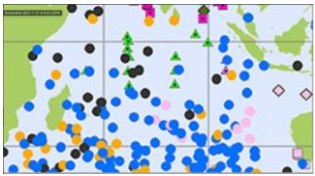
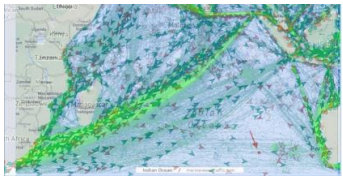

| **Fig 5: Screenshot of Moored observatories (Zoomed into Indian Ocean)** | **Fig 6: Merged Figures of 1 and 2** |

From the information gathered, we were able to merge and overlay Figs 1 and 2, in order to create mock-up of the map and paths taken by the ships. The merged Figures are as shown in Fig 6, and the simplified mock paths taken are as seen in purple lines in Fig 7.

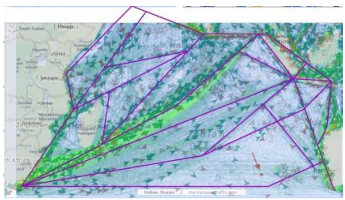
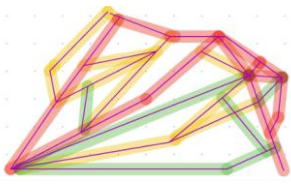

| **Fig 7: Simplified Ship Paths** | **Fig 8: Simplified Density Map** |

In order to translate the location of the Ocean Moors from Fig 5 to the mock-up map, the screenshot of moored observatories in Fig 5 was combined with the current mock-up in Fig 7. This is shown in Fig 9, where the red dots indicate the simplified locations of the Ocean Moors.

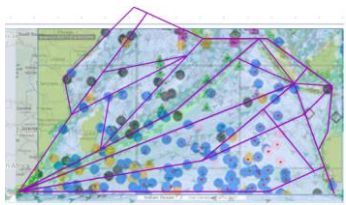

**Fig 9: Combined Figures with Ocean Moors**

Finally, the the map, together with the relative positions of the ocean moors are projected onto the original probable location on google maps in Fig 2 to create a clean backdrop for the ONE Simulator. This can be seen in Fig 10. In addition, the simplified paths with coastal guards are as indicated by the highlighted paths.

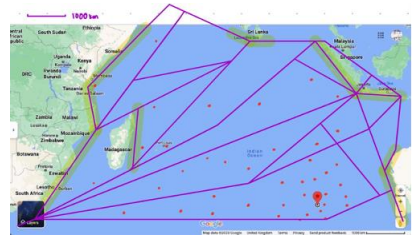

**Fig 10: Final Combined Map**

Using Fig 10, and a scale of 1:100km in ONE simulator, the coordinates of the corners and intersecting lines were found, and mapped to custom_map.wkt as shown in Fig 11.

```
# Lines (Paths)

LINESTRING (5000 2000, 0 0, 5000 2000)
LINESTRING (5000 2000, 30500 16000, 32000 31500, 21000 26000, 5000 2000)
LINESTRING (21000 26000, 18000 37000, 35000 54000, 40000 59000, 44000 57000, 30500 39000, 21000 26000)
LINESTRING (30500 39000, 44000 57000, 56500 44000, 30500 39000)
LINESTRING (21000 26000, 30500 39000, 56500 44000, 32000 31500, 21000 26000)
LINESTRING (30500 16000, 32000 31500, 56500 44000, 30500 16000)
LINESTRING (30500 16000, 61500 49500, 68000 49500, 78500 49500, 52500 27000, 30500 16000)
LINESTRING (5000 2000, 30500 16000, 52500 27000, 73000 45000, 95500 35500, 5000 2000)
LINESTRING (73000 45000, 78500 49500, 85500 35500, 73000 45000)
LINESTRING (78500 49500, 81500 49500, 93500 40500, 89500 35500, 78500 49500)
LINESTRING (5000 2000, 89500 35500, 61000 11500, 5000 2000)
LINESTRING (61000 11500, 80000 27500, 84500 22500, 61000 11500)
LINESTRING (80000 27500, 89500 35500, 93000 26000, 84500 22500, 80000 27500)
LINESTRING (89500 35500, 101000 35000, 93000 26000, 89500 35500)
#LINESTRING (5000 2000, 61000 11500, 84500 22500, 96000 8500, 81500 2000, 5000 2000)
LINESTRING (84500 22500, 93000 26000, 97000 15000, 99000 9900, 96000 8500, 84500 22500)
LINESTRING (93000 26000, 101000 35000, 100000 20500, 97000 15000, 93000 26000)
LINESTRING (99000 9900, 102000 1500, 99000 9900)

#LINESTRING (10000 45000, 20000 20000, 45000 10000, 10000 45000)
#LINESTRING (10000 5000, 20000 20000, 45000 40000, 10000 5000)

#LINESTRING (3 1, 1 59000, 5000 2000, 105000 59000, 105000 1, 1 1)
```

**Fig 11: custom_map.wkt**

Next, to analyze the density as shown in Fig 8, ships were split into 3 different groups (groups 2-4 as shown in Fig 16), each following different points of interests. The 3 different points of interest files are shipPOIs_1.wkt, shipPOIs_2.wkt, shipPOIs_3.wkt, which can be seen in Fig 12, 13 and 14 respectively.

```
POINT (5000 2000)
POINT (44000 57000)
#POINT (61500 49500)
POINT (78500 49500)
```

```
POINT (78500 49500)
POINT (93500 40500)
POINT (101000 35000)
POINT (97000 15000)
POINT (102000 1500)
POINT (89500 35500)
```

```
POINT (40000 59000)
POINT (18000 37000)
POINT (30500 39000)
POINT (56500 44000)
POINT (61000 11500)
POINT (73000 45000)
```

| **Fig 12: shipPOIs_1.wkt** | **Fig 13: shipPOIs_2.wkt** | **Fig 14: shipPOIs_3.wkt** |

*1)            Debris*

For the mobile nodes capable of wireless communication, such as electronic devices from passengers and crew, as well as potential debris, they will be re-named as plane crash debris. They follow MANETs, which would bring their communication range to 100m [10]. The range may be short, however, in this scenario, we will be simulating the dispersal of debris after the crash. According to a past case study of a similar crash (Air France Flight 447), on day 25, the debris scatter spanned 200 miles (320 kilometers) [11]. Considering the scaled down debris scatter to 4 days (time where families claimed phones continued to ring), the range of the communication of mobile nodes will be simplified and estimated to be 80km.

## 2) Ships

While ships use a variety of broadcasting interfaces, the broadcasting interface used by ships in this simulation will be Very High Frequency (VHF) as it is used for long range communication between ships. If a ship is surrounded by another within 30 km, a mesh/ad-hoc network can be created using VHF modem to build multi-hop network connections [12]. According to the average speed of vessels in the world merchant fleet in 2018 [13], ships ranging from general cargo ships and oil tankers to container ships and vehicle carrier, travel at average speeds between 9.25 – 14.95 knots ≈ 17 – 28km/h

## 3) Coastal Guards

In this simulation, coastal guards will follow the same broadcasting interface as general ships. While coast guard ships have top speeds ranging from 28 to 45 knots [14], we will assume that they patrol at 20knots – 28 knots ≈ 37 – 52 km/h, a generally faster speed than general ships, since the information is likely confidential.

## 4) Ocean Moors

For ocean surface communication of the ocean moors, as radio is traditionally used by moors to transmit data [15], we will assume that VHF is used and will share the same range as ships.

Using the collected information and providing rough estimates for the movement speeds and broadcast ranges, the details of each group can be seen in the table as seen in Fig 15.

For the buffer sizes of all groups, a size of 30MB has been chosen in order to avoid bottlenecks in the simulations.

| Group | Node Count | Movement Speed | Buffer Size | Broadcast Range |
|---|---|---|---|---|
| Plane Crash Debris | 5 | 0 | 30M | 80km |
| Ships | 170 | 17 – 28km/h | 30M | 30km |
| Coastal Guards | 20 | 37 – 52 km/h | 30M | 30km |
| Ocean Moors | 32 | 0 | 30M | 30km |

**Fig 15: Groups of Nodes in the Scenario**

### a) Distance

Transmission ranges in the simulation are in meters. We decided to use a scale of 1:100 in order to fit the map into the simulator.

For example, 30km = 30000m -> 300m (in ONE simulator)

### b) Time

For a speed of 28km/h, it can be approximated to 8m/s -> 0.08m/s (in ONE simulator) However, we are unable to key in 0.08 in ONE simulator. Hence, we used a time scale of 1:60 (1 second to 1 minute).

For example, 28km/h ≈ 8m/s -> 0.08m/s = 3meters/minute (in ONE simulator)

| Type | Groups | ID | Movement Speed | Interface Type | Broadcast Range |
|---|---|---|---|---|---|
| Plane Crash Debris | 1 | pcd | 0 | debrisInterface | 800 |
| Ships | 2-4 | s | 3 – 5 | VHFInterface, debrisInterface | 300 |
| Ocean Moors | 5-36 | om | 0 | VHFInterface | 300 |
| Coastal Guards | 37-41 | cg(af/ma/sr/in/au) | 6 - 8 | VHFInterface | 300 |

**Fig 16: Values Used for Each Group in ONE Simulator**

The table was converted to code changes in the configuration file as seen in Figs 17 – 21. In addition, the transmit speed of all interface types were set at 100M to avoid bottlenecks in the simulations.

```
Group1.groupID = Plane_Crash_Debris
Group1.nrofHosts = 5
Group1.movementModel = StationaryMovement
#Group1.movementModel = RandomWaypoint
#Group1.host1.initialPosition = (79500, 53000)
Group1.nodeLocation = 79500, 53000
Group1.bufferSize = 30M
Group1.nrofInterfaces = 1
Group1.interface1 = debrisInterface
```

**Fig 17: PCD Group**

```
Group2.groupID = s
Group2.nrofHosts = 50
Group2.movementModel= ShortestPathMapBasedMovement
Group2.speed = 3,5
Group2.bufferSize = 30M
Group2.nrofInterfaces = 1
Group2.interface1 = shipInterface
Group2.pois = 1, 1
PointsOfInterest.poiFile1 = data/shipPOIs_1.wkt
```

**Fig 18: S Group Example**

```
Group37.groupID = cgaf
Group37.nrofHosts = 7
Group37.movementModel= ShortestPathMapBasedMovement
Group37.speed = 6, 8
Group37.bufferSize = 30M
Group37.nrofInterfaces = 1
Group37.interface1 = shipInterface
Group37.pois = 4, 1
PointsOfInterest.poiFile4 = data/coastalGuardAfricaPOIs.wkt
```

**Fig 19: CG Group Example**

```
Group5.groupID = om
Group5.nrofHosts = 1
Group5.movementModel = StationaryMovement
Group5.nodeLocation = 29000, 55000
Group5.bufferSize = 30M
Group5.nrofInterfaces = 1
Group5.interface1 = mooringInterface
```

**Fig 20: OM Group Example**

```
## Interface-specific settings:
debrisInterface.type = SimpleBroadcastInterface
debrisInterface.transmitSpeed = 100M
debrisInterface.transmitRange = 800

VHFInterface.type = SimpleBroadcastInterface
VHFInterface.transmitSpeed = 100M
VHFInterface.transmitRange = 300
```

**Fig 21: Interface Settings**

While all code changes will be seen in the configuration file, these are the important components that contribute to the final output of the simulator as seen in Fig 22.

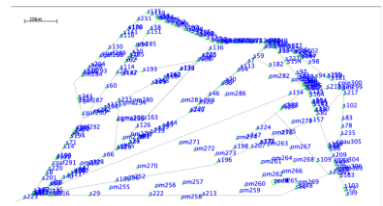

**Fig 22: Output of Simulator**

### IV. RESULTS

Upon initial inspection, and as seen from Fig 22, it is evident that the range of communication for the plane crash debris is insufficient to reach other ships. As seen from Fig 23, which was a screenshot taken at the end of the simulation, the debris could only make connections within themselves, showing an inability to establish links with other ships. Hence, this prevented the transmission of messages to emergency services.

```
  4 connection(s)
    pcd0<->pcd4
    pcd0<->pcd1
    pcd0<->pcd3
    pcd2<->pcd0
```

**Fig 23: pcd Connections**

The simulation outcome strongly suggests that the technology and capabilities of modern communication networks are insufficient to facilitate the transmission of emergency information and whereabouts of the crash to emergency services or even other ships. This finding provides insight into the challenges associated with delivering crucial information in a remote plane crash, potentially explaining the prolonged unknown location of the aircraft over the past decade.

The inherent constraint of limited network range attributed to the mobile nodes within the plane debris poses a significant challenge. Despite extending the communication ranges of ships, messages cannot be sent due to the ships being outside of the sender's (mobile nodes) range. Unless the mobile nodes can cover vast distances of hundreds of kilometres, or that there exists an extensive network of ocean moors spanning the ocean surface, the reception of messages will remain unattainable. Nevertheless, the focus will shift towards optimising the likelihood of successful message delivery in scenarios where ships do manage to receive a remote emergency message.

Therefore, despite the initial limitations of the communication range, we have extended the project by shifting the plane crash location to enable communication with the closest ship. This extension will allow for the simulation and comparison of different protocols in scenarios where ships are within range to receive emergency messages. It enables a more comprehensive assessment of the suitability of the different protocols for a remote emergency and will be discussed in the parts following this. The changes are as seen in Figs 24 and 25.

```
## Interface-specific settings:
debrisInterface.type = SimpleBroadcastInterface
debrisInterface.transmitSpeed = 100M
debrisInterface.transmitRange = 10
```

```
Group1.groupID = pcd
Group1.nrofHosts = 5
Group1.movementModel = StationaryMovement
Group1.nodeLocation = 79500, 57000
Group1.bufferSize = 30M
Group1.nrofInterfaces = 1
Group1.interface1 = VHFInterface
```

**Fig 24: Changes to transmitRange**   **Fig 25: Changes to nodeLocation**

Next, to evaluate the performances of each protocol, we will be focusing on mainly 2 key metrics, average latency and delivery probability, as they contribute directly to the effectiveness of communication systems during critical systems.

Average latency is information on the time taken for messages to be successfully delivered, which is a critical factor in emergency scenarios where effective communication is crucial. Simultaneously, delivery probability is a key factor in indicating that emergency messages reach the intended recipients (coastal guards).

With all the information and changes to the configuration thus far, we conducted a simulation spanning a critical period of four days, a timeframe pivotal for emergency response, until the point as mentioned in section 2.1 where the mobile nodes were claimed to stop working. The outcomes of the simulation using maxProp and Epidemic routing are as seen in Figs 26 and 27 respectively.

```
sim_time: 5760.0000
created: 194
started: 2674
relayed: 2674
aborted: 0
dropped: 2315
removed: 11
delivered: 4
delivery_prob: 0.0206
response_prob: 0.0000
overhead_ratio: 667.5000
latency_avg: 4691.8000
latency_med: 4929.5000
hopcount_avg: 3.0000
hopcount_med: 3
buffertime_avg: 233.6528
buffertime_med: 5.4000
rtt_avg: NaN
rtt_med: NaN
```

```
sim_time: 5760.0000
created: 194
started: 43367
relayed: 43365
aborted: 0
dropped: 43007
removed: 0
delivered: 4
delivery_prob: 0.0206
response_prob: 0.0000
overhead_ratio: 18840.2500
latency_avg: 4691.8000
latency_med: 4929.5000
hopcount_avg: 3.0000
hopcount_med: 3
buffertime_avg: 26.0341
buffertime_med: 4.0000
rtt_avg: NaN
rtt_med: NaN
```

**Fig 26: MaxProp Routing**   **Fig 27: Epidemic Routing**

As seen from the simulation results of the two different protocols, it can be seen that, at the current technological state, both routing protocols produce nearly identical results. This is particularly evident in the two key metrics, delivery probability and average latency, which demonstrate identical values.

At this juncture, discerning the superior protocol is challenging considering the similar results. However, considering the continuous advancements in technology, it is a reasonable assumption that communication ranges may experience growth. To explore the potential implications of these advancements, we can alter the variables of the simulation. This would allow us to ascertain if a protocol outperforms the other with increasing ranges, or even discern the optimal communication range that technology could aim towards to ensure the highest probability of delivery.

As the initial range is set at 300km, subsequent simulations will be conducted with increments of 300km, continuing until it reaches 3900km. This span represents slightly over ten times the original distance, providing sufficient data for meaningful comparisons. The results for both protocols are as seen in Figs 28 and 29.

| range | created | started | relayed | aborted | dropped | removed | delivered | delivery_pro | response_pr | overhead_ra | latency_avg | latency_med | hopcount_a | hopcount_m | buffertime_a | buffertime_med |
|---|---|---|---|---|---|---|---|---|---|---|---|---|---|---|---|---|
| 300 | 194 | 2636 | 2636 | 0 | 2276 | 11 | 4 | 0.0206 | 0 | 658 | 4691.8 | 4929.5 | 3 | 3 | 234.7139 | 4.3 |
| 600 | 194 | 2791 | 2791 | 0 | 2345 | 17 | 5 | 0.0258 | 0 | 557.2 | 4715.28 | 4892.3 | 4 | 4 | 235.2936 | 6.5 |
| 900 | 194 | 2899 | 2899 | 0 | 2390 | 38 | 7 | 0.0361 | 0 | 413.1429 | 4553.8857 | 4435.9 | 3 | 3 | 227.0165 | 5.7 |
| 1200 | 194 | 3808 | 3808 | 0 | 3130 | 63 | 9 | 0.0464 | 0 | 422.1111 | 4359.6778 | 4303.2 | 4.1111 | 4 | 196.9212 | 2.1 |
| 1500 | 194 | 3908 | 3908 | 0 | 3057 | 90 | 7 | 0.0361 | 0 | 557.2857 | 4245.9143 | 4193.2 | 7.1429 | 7 | 221.2 | 9 |
| 1800 | 194 | 4365 | 4365 | 0 | 3466 | 87 | 7 | 0.0361 | 0 | 622.5714 | 4126.3 | 4114.8 | 7.4286 | 7 | 214.0889 | 24.6 |
| 2100 | 194 | 5873 | 5873 | 0 | 4783 | 224 | 12 | 0.0619 | 0 | 488.4167 | 3993.025 | 4020.3 | 6.4167 | 7 | 178.069 | 11.7 |
| 2400 | 194 | 6819 | 6819 | 0 | 5593 | 357 | 14 | 0.0722 | 0 | 486.0714 | 3777.4429 | 3708.2 | 6.1429 | 6 | 163.9438 | 7.3 |
| 2700 | 194 | 6846 | 6846 | 0 | 5595 | 375 | 13 | 0.067 | 0 | 525.6154 | 3703.7769 | 3691.8 | 6.6154 | 6 | 159.4989 | 14 |
| 3000 | 194 | 8924 | 8924 | 0 | 6078 | 772 | 22 | 0.1134 | 0 | 404.6364 | 3751.1227 | 3582.5 | 9.0909 | 7 | 157.7913 | 7.2 |
| 3300 | 194 | 10026 | 10026 | 0 | 7025 | 800 | 22 | 0.1134 | 0 | 454.7273 | 4058.3227 | 4116.8 | 10 | 8 | 144.9494 | 4.3 |
| 3600 | 194 | 11278 | 11278 | 0 | 8054 | 898 | 23 | 0.1186 | 0 | 489.3478 | 3684.7609 | 3418.3 | 8.2609 | 6 | 165.4131 | 3 |
| 3900 | 194 | 14015 | 14014 | 0 | 8362 | 2087 | 29 | 0.1495 | 0 | 482.2414 | 3600.1207 | 3493 | 12.2759 | 11 | 163.6565 | 12.6 |

**Fig 28: MaxProp Simulation Data**

| range | created | started | relayed | aborted | dropped | removed | delivered | delivery_pro | response_pr | overhead_ra | latency_avg | latency_med | hopcount_a | hopcount_m | buffertime_a | buffertime_med |
|---|---|---|---|---|---|---|---|---|---|---|---|---|---|---|---|---|
| 300 | 194 | 43367 | 43365 | 0 | 43007 | 0 | 4 | 0.0206 | 0 | 10840.25 | 4691.8 | 4929.5 | 3 | 3 | 26.0341 | 4 |
| 600 | 194 | 40924 | 40922 | 0 | 40463 | 0 | 5 | 0.0258 | 0 | 8183.4 | 4715.28 | 4892.3 | 4 | 4 | 27.5448 | 4.1 |
| 900 | 194 | 27277 | 27276 | 0 | 26777 | 0 | 5 | 0.0258 | 0 | 5454.2 | 4651.68 | 4719.5 | 3 | 3 | 40.0629 | 4.1 |
| 1200 | 194 | 58391 | 58389 | 0 | 57788 | 0 | 9 | 0.0309 | 0 | 9730.5 | 4456.7833 | 4648.5 | 9 | 7 | 18.4598 | 3.9 |
| 1500 | 194 | 54731 | 54730 | 0 | 53940 | 0 | 5 | 0.0258 | 0 | 10945 | 4327.8 | 4279.2 | 8.2 | 8 | 20.4911 | 4 |
| 1800 | 194 | 89284 | 89282 | 0 | 88502 | 0 | 5 | 0.0258 | 0 | 17855.4 | 4048.26 | 4045.6 | 27.4 | 26 | 10.9996 | 4 |
| 2100 | 194 | 65545 | 65544 | 0 | 64634 | 0 | 9 | 0.0464 | 0 | 7281.6667 | 4111.0556 | 4086.9 | 34.7778 | 35 | 17.0277 | 4 |
| 2400 | 194 | 56881 | 56879 | 0 | 55969 | 0 | 8 | 0.0412 | 0 | 7108.875 | 3753.9875 | 3754.8 | 22.5 | 26 | 19.5884 | 4.1 |
| 2700 | 194 | 78882 | 78880 | 0 | 77971 | 0 | 9 | 0.0464 | 0 | 8763.4444 | 3778.8556 | 3704.2 | 25.8889 | 30 | 14.1768 | 4.1 |
| 3000 | 194 | 66587 | 66585 | 0 | 64485 | 0 | 12 | 0.0619 | 0 | 5547.75 | 3948.95 | 4155.9 | 30.3333 | 31 | 15.0441 | 3.9 |
| 3300 | 194 | 37969 | 37969 | 0 | 35566 | 0 | 15 | 0.0773 | 0 | 2530.2667 | 4412.7733 | 4613.1 | 49.6 | 50 | 31.3505 | 4.2 |
| 3600 | 194 | 70084 | 70082 | 0 | 67574 | 0 | 16 | 0.0825 | 0 | 4379.125 | 3878.9875 | 4289.9 | 54.5 | 56 | 17.6219 | 4.1 |
| 3900 | 194 | 83346 | 83344 | 0 | 79599 | 0 | 23 | 0.1186 | 0 | 3622.6522 | 3956.8739 | 4326.4 | 98.3043 | 105 | 15.2195 | 4.1 |

**Fig 29: Epidemic Simulation Data**

Taking a look at the two key metrics, a graph of delivery probability and latency average against range is created as shown in Figs 30-31 and 32-33 respectively. In order to visualise the data better, using a second order polynomial to create a trend line, a curve of best fit was also added to Figs 30 and 31, while connecting lines between data points are added to Figs 32 and 33.

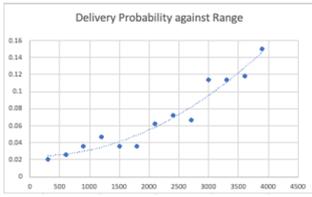

**Fig 30: Delivery Probability Graph (MaxProp)**

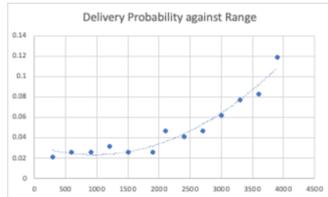

**Fig 31: Latency Average Graph (Epidemic)**

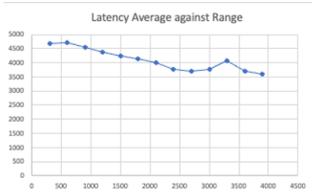

**Fig 32: Latency Average Graph (MaxProp)**

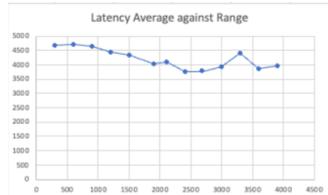

**Fig 33: Latency Average Graph (Epidemic)**

Upon initial examination, it is evident that both routing protocols take advantage of the extended communication distances, as shown by the increasing delivery probability and reduction in latency.

A closer inspection reveals an interesting trend however: the delivery probability exhibits a rate of increase surpassing that of the range increment. This seems to suggest as the communication range increases, both protocols utilise the network more efficiently.

Furthermore, both the graph of latency average show a notable spike at range 3300km despite the downward trend. It is possible that the anomaly is an indication of a transition point where the extended range starts to introduce other problems, possibly related to the larger communication distances as a result of the increase in network complexity.

Moreover, at approximately 2500km on both latency average graphs, a local minimum can be observed at that range. This observation suggests that this range could represent an optimal point where the protocols utilise the network more effectively.

To make better comparisons between the protocols, the data points of delivery probability and latency average were plotted on the same graph as seen in Figs 34 and 35 respectively.

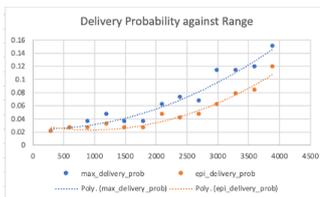

**Fig 34: Delivery Probability Graph (Combined)**

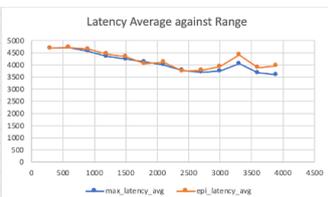

**Fig 35: Latency Average Graph (Combined)**

From Fig 34, it is evident that despite both protocols beginning with identical delivery probabilities, MaxProp protocol not only consistently maintains a higher delivery probability, but also exhibits a higher rate of increase.

Furthermore, in Fig 35, although both protocols exhibit a similar average latency initially, MaxProp protocol begins to display superior performance beyond the point of local minimum as mentioned earlier, consistently maintaining a lower average latency.

In addition, while the latency average of MaxProp protocol decreases towards the end of the data, the latency average of Epidemic protocol sees a rise. This seems to suggest that there potential for MaxProp protocol to utilise the network more efficiently, whereas Epidemic protocol may be approaching its limit for further improvement.

### V. SUMMARY OF RESULTS AND CONCLUSION

The initial evaluation of the simulation suggested that, given the current state of technology, it was unlikely that distress messages could have been successfully delivered, as seen by the inability to establish communications with other nodes within the original simulated ranges.

Recognising these challenges, the project scope was extended by adjusting the plane crash location to enable communication with the closest ship. Upon doing so, the performance of MaxProp and Epidemic protocols were examined under a larger hypothetical range, recognising its relevance as technology is continuously advancing.

Focusing on the two key metrics of delivery probability and average latency, MaxProp consistently outperformed Epidemic and displayed a notable potential for further improvement. The observation seems to suggest that MaxProp may emerge as a more suitable protocol as technology advances and communication ranges become larger.

In addition, a study of both protocol performances suggest the presence of a local minimum when the range is approximately 2700km. While the exact value may vary, it hints at the possibility that, as communication range increases, there may exist an optimal point where protocols utilise the network more effectively.

The simulation offers both advantages and disadvantages in modelling the communication networks in a remote sea emergency scenario. It is advantageous as it serves as a good visual tool to provide a clear depiction of the communication range of nodes in the sea. This facilitates a better understanding of ship ranges, mobile node capabilities, and the potential locations of plane debris after a crash, in order to assess the communication possibilities.

However, a disadvantage would be the reliance of an overly simplified 2D map. Due to this, the model not only neglects the diverse paths that ships may take on the ocean surface, but also the paths that other vehicles with communication capabilities such as submarines or satellites take. This lack of representation diminishes the completeness of the communication network model. Moreover, the simulation employs fixed communication ranges which does not account for the dynamic real-world factors such as weather conditions and the distance between nodes, reducing the realism of the model.

## VI. DISCUSSIONS

In considering potential enhancement to the simulation, there may be several areas for improvement. Firstly, it may be a good option to explore alternative simulators with 3D space capabilities such as NS3 [16], potentially increasing the realism of the simulation by the incorporating underwater and air traffic. Additionally, the implementation of a more detailed map with more ship routes may also contribute to more accurate portrayal of the maritime scenarios allowing for a more accurate representation of sea traffic densities.

In conclusion, the simulations and evaluations presented valuable insights into the performance of routing protocols in a remote sea emergency scenario. While the simulation suggests that MaxProp may be more suitable with increasing communication ranges in a remote sea emergency, it also shows the need for continuous technological advancements to enhance communication ranges and improve the efficiency of protocols in such critical situations.

Given the compelling indications from the original simulation that it is not possible to successfully deliver an emergency message successfully even with increased ranges, there arises a need for further exploration of other alternatives. It may be good to consider using long range acoustic communication [17] between ocean mooring communication nodes which could have the potential to be a game-changer, not only making it possible to locate remote crash debris but also greatly increase the probability of message delivery in a remote sea emergency.